\documentclass[prb,twocolumn,preprintnumbers,nofootinbib,floatfix]{revtex4-1}
\usepackage{amssymb}
\usepackage{amsmath}
\usepackage[english]{babel}
\usepackage[latin1]{inputenc}
\usepackage{bbm,bm,slashed}
\usepackage{graphicx}
\usepackage{color,array,tabularx}

\graphicspath{ {./images/} }

\begin{document}

\title{Functional renormalization group study of an eight-band model for the
iron arsenides }
\date{July 9, 2014}
\author{J. Lichtenstein$^1$, S. A. Maier$^1$, C. Honerkamp,$^1$, C. Platt$^2$%
, R. Thomale$^{2}$, O.K. Andersen$^3$ and L. Boeri$^{3,4}$}

\affiliation{$^1$ Institute for Theoretical Solid State Physics, RWTH Aachen University, D-52056 Aachen, Germany  \\  and JARA - FIT Fundamentals of Future Information Technology 
}

\affiliation{$^2$ Institute for Theoretical Physics and Astrophysics, University of W\"urzburg, D-97074
W\"urzburg, Germany}

\affiliation{$^3$ Max-Planck-Institute for Solid State Research, Stuttgart,
D-70569 Stuttgart, Germany}

\affiliation{$^4$ Institute for Theoretical Physics,  Graz University of
Technology, 8010 Graz, Austria}

\begin{abstract}
We investigate the superconducting pairing instabilities of eight-band
models for the iron arsenides. Using a functional renormalization group
treatment, we determine how the critical energy scale for superconductivity
depends on the electronic band structure. Most importantly, if we vary the
parameters from values corresponding to LaFeAsO to SmFeAsO, the pairing
scale is strongly enhanced, in accordance with the experimental observation.
We analyze the reasons for this trend and compare the results of the
eight-band approach to those found using five-band models.
\end{abstract}

\maketitle

\section{Introduction}

Superconductivity in iron arsenides has been a central field of studies in
contemporary solid state physics\cite%
{Paglione2010,Stewart2011,Hirschfeld2011,Scalapino2012,PlattReview}. While many
interesting relations and effects have been uncovered, there is still only
rudimentary understanding of what should be done to achieve higher
transition temperatures exceeding the 55K of SmFeAsO\cite{55K}. One of the
reasons is the complexity of these materials. While DFT techniques can be
used to understand the lattice and electronic structure to quite some
precision\cite{boeri}, our tools for the computation of electronically
mediated pairing are still incomplete and will need more refinement to reach
predictive power. Furthermore, the many-body calculations that have been
performed for these multi-band systems exhibit a rather strong dependence on
details of the microscopic Hamiltonian and also on the level of
approximations. For example, the dominant pairing state, in particular
the competition between $d$-wave and $s^\pm$-wave, sensitively depends on the model parameters\cite{kuroki,graser,PhysRevB.80.140515,platt_sid,ThomaleMay2011,ThomaleSep2011,PhysRevB.85.180502}. While this can
reflect the physical reality, one should aim for an understanding of how
robust the theoretical results are with respect to the uncertainties of the
underlying model and the theoretical approach. This implies testing
experimental materials trends against the theoretical picture and also
comparing whether different theoretical models arrive at comparable results.

Regarding material trends, a key issue is the systematics of the
superconducting $T_{c}$ with the system parameters. Here, the pnictogen
height, i.e. the height of the As atom above or below the Fe-planes in FeAs
compounds, has been recognized as an important tuning parameter. This was
argued by Kuroki et al.\cite{Kuroki2009} in an influential paper already in
2009. These authors showed that the pairing eigenvalues obtained in an RPA
treatment of the pairing interaction grow when the pnictogen height is
increased. In their 5-band model, the band having iron $d_{X^{2}-Y^{2}}$ $%
\left( \equiv d_{xy}\right) $ character\footnote{
    Kuroki et al. \cite{Kuroki2009} have the $d_{xy}$ pocket at $(\pi ,\pi )$ while we have it at 
    $(0,0)$. Similarly, Kuroki et al. have the two other hole pockets at $(0,0)$ while we have 
    them at $(\pi ,\pi )$. The reason is given in Fig. 3 of Ref \onlinecite{AndersenBoeri2011}.
    In general, the pseudo-momentum used by Kuroki et al. differs by $(\pi ,\pi )$
    from our $\mathbf{k}$-vector which enumerates the irreducible representations of the
    Abelian group generated by the glide-mirror operation \cite{LeeWen2008}.
} undergoes strong changes with pnictogen height. When the
height is as low as in LaOFeP, the $xy$ hole pocket ceases to exists and is
replaced by a $d_{3z^{2}-1}$ pocket. The disappearance of the $d_{xy}$
pocket causes the system to switch from a high-scale pairing phase with a
nodeless gap function to a low-scale pairing phase with a very anisotropic
gap function, possibly with nodes on the electron pockets. These findings%
\cite{Kuroki2009} have also been confirmed by other groups, e.g. by using
FLEX\cite{ikeda} or the functional RG\cite{ThomaleMay2011}. For the
latter, a consistent picture was provided not only for the
superconducting state, but also the change of the magnetic fluctuation profile.


Another possibility to describe the electronic structure of the iron
pnictides, is to use an 8-band model which in addition to the 5 iron $d$
orbitals includes 3 pnictide $p$ orbitals. Such a model was worked out by
Andersen and Boeri\cite{AndersenBoeri2011}. Whereas in the 5-band
description the pnictide $p$-characters are folded into the tails of the $d$%
-like Wannier orbitals, which thereby become less localized and less $d$%
-like, the Wannier orbitals in the 8-band description are fairly localized
and atomic-like. With that description, one can explicitly see that
decreasing the pnictogen height, increases the hopping integral, $t_{xy,z},$
between the pnictide $p_{z}$ and iron $d_{xy}$ orbitals and thereby moves
the $pd$ anti-bonding band \emph{up}wards. Together with this band, the $%
d_{xy}/p_{z}$-like parts of the electron pockets move and, hence, the Fermi
level. Eventually (by LaOFeP) the Fermi level is \emph{above} the top of the 
$d_{xy}$ hole pocket which --by being placed at a high-symmetry point--
cannot hybridize with the pnictide $p_{z}$ orbital and therefore does not
move (with respect to the one-electron potential)\cite{AndersenBoeri2011}.

In the present paper we shall see that although for weakly and optimally
electron-doped iron \emph{arsenides} the $d_{xy}$ hole pocket always exists,
the \emph{de}crease of $p_{z}$-$d_{xy}$ hybridization caused by an 
\emph{in}crease of the As height (e.g. SmOFeAs) leads to an \emph{in}crease of the
pairing scale. 
This means that the experimentally observed trend may be understood as a change of a single band structure parameter that is most transparently implemented in an 8-band description.
More precisely, we shall try to understand the difference
between the pairing-energy scales for LaFeAsO, with an experimental\cite%
{kamihara} $T_{c}$ of $\lesssim 30K,$ and SmFeAsO\cite{55K} with $T_{c}\sim
55K,$ i.e. roughly twice as high. These two systems have the same layered,
so-called 1111 structure, but the pnictogen height is larger --and the Fe-As
tetrahedra closer to being regular-- in SmOFeAs than in LaOFeAs. In
addition, we shall study the effect of loosing the $d_{xy}$ pocket, as it
happens upon overdoping these arsenides.

We do not claim that this more refined effect of the pnictogen height within
the five-pocket regime cannot be found in a 5-band description, it is just
that it can be obtained very transparently in the 8-band description. Besides
exploring the consequences of the variation in pnictogen height, our present
8-band study is also interesting from a methodological point of view.
Describing the band structure over a larger energy window should a priori
not change the theoretical predictions, but due to the approximations made
for the Coulomb correlations, the outcome of 8-band and 5-band calculations
can indeed be different. In the usual approximation used for the 5-band
model, only the local Coulomb repulsion between the effective Fe $d$
orbitals is considered in terms of density-density and Hund's rule spin-spin
interactions. In $\left( \mathbf{k,}\omega \right) $-orbital-space, this corresponds
to a constant interaction. In the 8-band model, however, it is natural to keep also the Coulomb
repulsion $U_{pd}$ between Fe $d$ and As $p$ orbitals, thus leading to $%
\mathbf{k}$-dependent interactions already at the bare level.
It may also be worth pointing out, that whereas in the 5-band
description, where only the $pd$ \emph{anti}bonding $d$ characters are
present, the configuration is $p^{6}d^{6},$ it is $p^{4.5}d^{7.5}$ in the
8-band DFT description\cite{Schickling}\cite{Haverkort}.

Our theoretical approach will be the fRG (functional renormalization group;
for recent reviews, see Refs. \onlinecite{metznerRMP,PlattReview}). Such fRG studies have recently
been applied to a number of 5-band scenarios\cite{}. The fRG exhibits
similar systematics for the pairing states as the commonly-used RPA and FLEX
(fluctuation-exchange approximation)\cite{kuroki,graser,ikeda} approaches,
but in addition gives reasonable information about the energy scale of the
pairing instabilities. In particular, this ansatz allows for a
combined DFT-fRG approach, as has been successfully pioneered in the
context of LiFeAs where the $s^\pm$ SC order prevails as in many other
iron-based superconductors despite a significantly deviating magnetic
fluctuation profile~\cite{PhysRevB.84.235121}. 
The fRG sums all one-loop diagrams to infinite order in the \emph{bare}
interaction. As a subset, it contains the usual ladder sums and RPA
summations. These can be understood as geometric series.
Hence, the fRG is not
biased in favor of one particular kind of ordering. This allows one to
obtain tentative phase diagrams having regions with leading
antiferromagnetic order bordering superconducting regions.

Here, we shall present a first fRG study of an 8-band model for the iron
arsenides. The paper is organized as follows: In Sec. \ref{sec:model} we
describe our model. In Sec. \ref{sec:method} we give more details about our
fRG calculation used to search for superconducting pairing. In Sec. \ref%
{sec:La1111} we present the relevant fRG results for model parameters
corresponding to LaOFeAs, and in Sec. \ref{sec:Sm1111} we show the
corresponding data for model parameters adequate for SmOFeAs. In Sec. \ref%
{sec:trend} we exhibit and discuss the material trend in terms of the
pairing scale when the model parameters are varied continuously from the La-
to the Sm situation. We conclude with a discussion of these findings in Sec. %
\ref{sec:conclusions}.

\section{Model}

\label{sec:model} The starting point for our investigations is the band
structure obtained ab-initio\cite{AndersenBoeri2011} by DFT-GGA calculations for LaOFeAs and
presented in the form of the hopping integrals between --and the on-site
energies of the 5 Fe $d$ and 3 As $p$ Wannier orbitals (NMTOs)\footnote{The 5-set (and the 8-set) Wannier orbitals are generated in the tetragonal $%
XY$-system and then linear combined to the cubic system, e.g.: $%
d_{xz}=\left( d_{Xz}-d_{Yz}\right) /\sqrt{2}.$ Due to the tails on As, those
linear combinations are not simply 45$^{\circ}$ rotations (see Fig. 5 in Ref. \onlinecite{AndersenBoeri2011}).}. As will be discussed later, we also use ab-initio
derived Coulomb interaction parameters. The energy bands with substantial La
or O character are more than 2 eV from the Fermi level and are not spanned
by our 8-orbital basis set. Only their hybridization with the 8 bands near
the Fermi level are included (via the La and O characters downfolded into
the tails of the 8 orbitals).

The so-called 1111 structure of LnOFeAs consists of alternating FeAs and LnO
layers. The FeAs layer has Fe at the corners of a square lattice with
primitive translations $a\mathbf{x}$ and $a\mathbf{y},$ and As alternatively
above and below the centers of the squares\footnote{%
The ``cubic'' $x$ and $y$ axes point from Fe to the nearest Fe neighbors on
the square lattice, while the ``tetragonal'' $X$ and $Y$ axes are turned by 45$%
^{\circ}$ and point towards the projection of the pnictide onto the Fe plane. $%
z=Z.$}. The LnO layer has the same
structure, but with O substituted for Fe and Ln for As. Due to the
above-below alternation of As and Ln, the primitive cell of the translation
group contains \emph{two} LnOFeAs units. However, a \emph{single} FeAs (or
LnOFeAs) layer can be generated by an Abelian group whose elementary
operation is a translation $a\mathbf{x}$ or $a\mathbf{y,}$ combined with the 
$z\rightarrow -z$ reversal ('move one step and stand on your head')\cite%
{AndersenBoeri2011}. If one lets the 2D Bloch vector, $\mathbf{k,}$ label
the irreducible representations of that group, then the primitive cell
contains only one formula unit and the Brillouin zone is a \emph{large}
square with edge $2\pi /a.$ Hopping \emph{between} FeAs layers, introduces
coupling between $\mathbf{k}$ and $\mathbf{k}+\left( \pi ,\pi \right) /a$
whereby the BZ is folded into the usual \emph{small} one, and is given a
height $2\pi /c.$ In the LnOFeAs structure the interlayer hopping is mainly
between the neighboring As $z$ orbitals and is small for the bands near the
Fermi level\cite{AndersenBoeri2011}. In the present work, interlayer
coupling will be neglected, although this is better justified for LaOFeAs
than for SmOFeAs, as may be seen from Fig.$\,$9 in Ref.$\,$%
\onlinecite{AndersenBoeri2011}.

\begin{figure}[t]
\includegraphics[width=0.23\textwidth]{spaghetti_paper.eps} %
\hspace{1.6mm}
\includegraphics[width=0.23\textwidth]{spaghetti_sm_paper.eps} %
\includegraphics[width=0.23\textwidth]{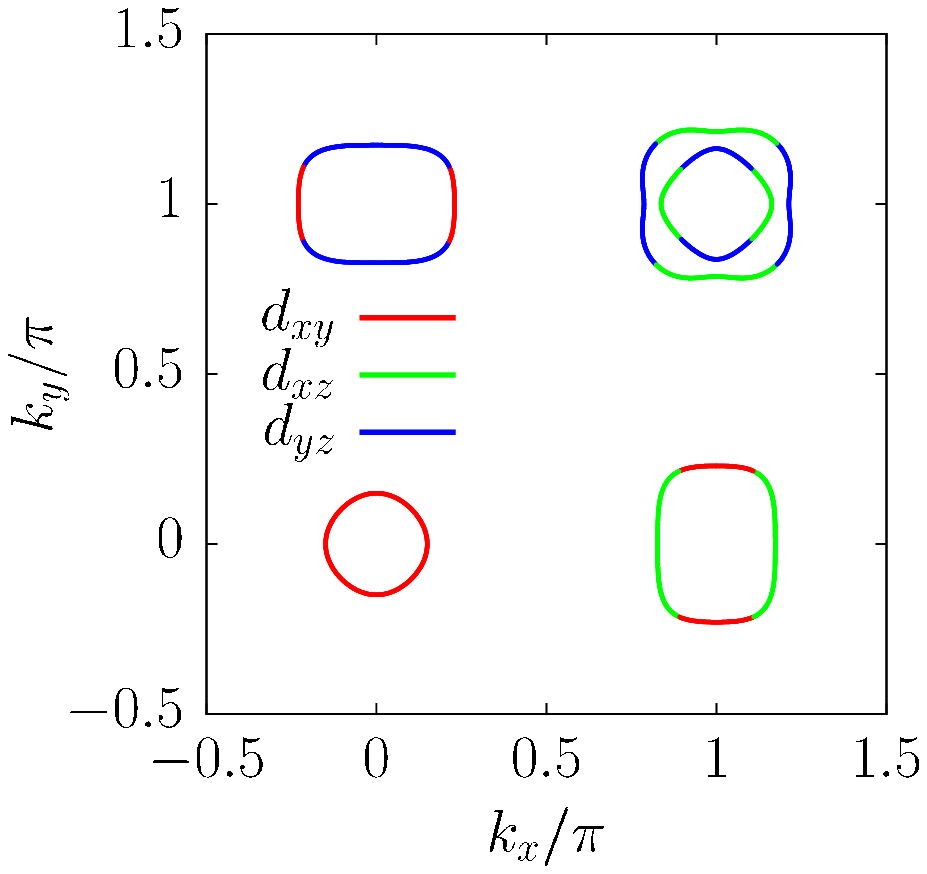} %
\includegraphics[width=0.23\textwidth]{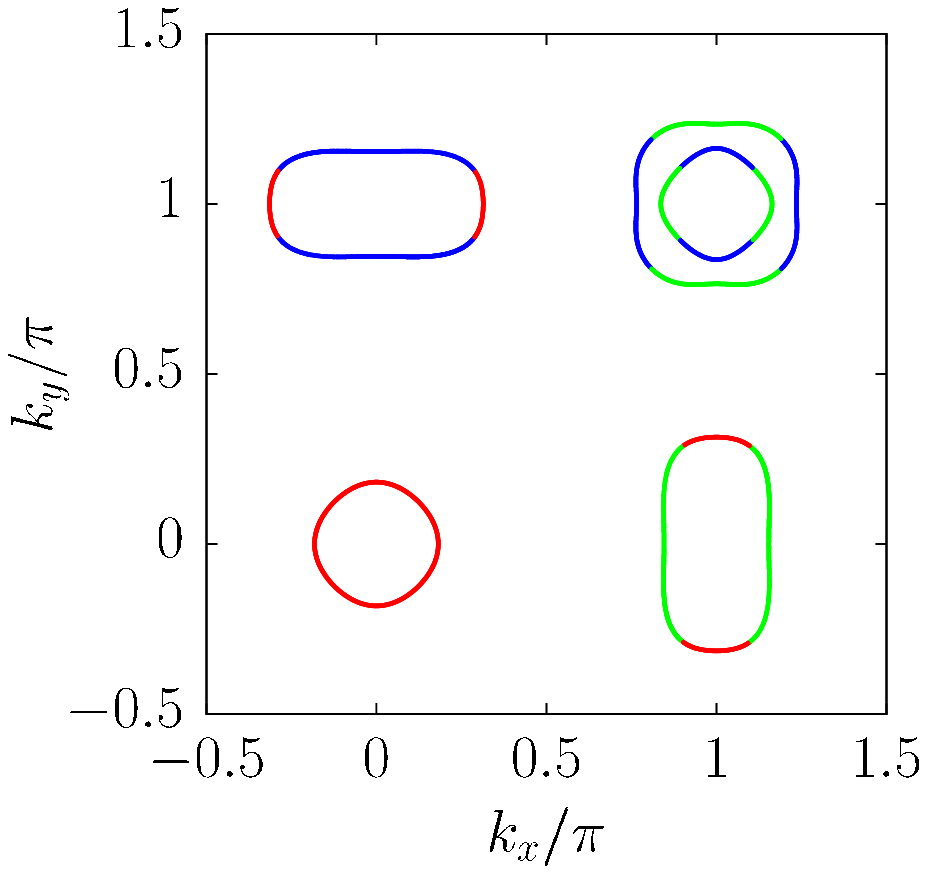}
\caption{Upper panels: Band structures of La-1111 (left) and Sm-1111 (right)
from the 8-band Hamiltonian introduced in the text. On top of the La bands
(black) we have plotted the Sm bands (red dashes). Near \={M}, at positive
energies, the uppermost, $p_{z}$-dominated band of the La band structure
moves down in energy and becomes the second-highest band for Sm. Lower
panels: Fermi surfaces (La left, Sm right) for the undoped compounds. The
color gives the dominant orbital character.}
\label{fig:models}
\end{figure}

The DFT-GGA-NMTO one-electron Hamiltonian in site and orbital representation
is:%
\begin{equation}
H_{0}=\sum_{i,j}\sum_{{o,o^{\prime }}\atop{s}}c_{i,o,s}^{\dagger
}h_{ij}^{oo^{\prime }}c_{j,o^{\prime },s}\,,  \label{BS}
\end{equation}%
where the $i$ and $j$ run over the square lattice, and $o$ and $o^{\prime }$
over the 8 orbitals. The matrix element $h_{ij^{\prime }}^{oo^{\prime }}$ is
the on-site energy of an orbital if $i=j$ and $o=o^{\prime },$ and a hopping
integral (or a crystal-field term) if $o\neq o^{\prime }$ or $i\neq j.$
Projecting the Hamiltonian into $\mathbf{k}$-space and diagonalizing the 8$%
\times $8 matrices for each $\mathbf{k,}$ results in 8 bands and
eigenvectors. The overall bandwidth of this 8-band complex is $\sim 7$eV, as
can be read off from Fig. \ref{fig:models}. In order to use this model for
electron-doped LaO$_{1-x}$F$_{x}$FeAs, we use the rigid band approximation
according to which the chemical potential, $\mu ,$ is shifted so far upwards
with respect to the bands that the growth of the electron pockets plus the
shrinkage of the hole pockets amount to $x/2$ times the area of the BZ.

Now, let us discuss the fermiology in more detail (see also Sect. 3.3 in Ref.%
$\,$\onlinecite{AndersenBoeri2011}). Within the range of relevant dopings,
the $e_{g}$-like $\left( d_{x^{2}-y^{2}},d_{3z^{2}-1}\right) $ bands are
gapped around the Fermi level, which is therefore crossed by only the $%
t_{2g} $-like bands. Of these, the $d_{xy}$ band is pure at $\bar{\Gamma}$ $%
\left( 0,0\right) $ --i.e. not hybridizing with any of the other orbitals in
the basis set. 
Because of the glide-mirror symmetry of the Fe-As lattice, the $d_{xy}$ Bloch wave at $k=0$ is out-of-phase on nearest-neighbor Fe sites,          
i.e. it has anti bonding $dd\pi$ character.
Therefore, it forms the top of the band.
With the Fermi level a bit lower, this band gives
rise to a $\bar{\Gamma}$-centered, nearly circular hole pocket which can be
seen in the lower plots of Fig. \ref{fig:models}.

Similarly, the $d_{xz}$ and $d_{yz}$ bands are nearly pure at \={M} $\left(
\pi ,\pi \right) ,$ where they are degenerate and antibonding $dd\pi .$ The
top of the $d_{xz}$ and $d_{yz}$ bands at \={M} are pushed up by weak
hybridization with As $p_{x}$ and $p_{y}$ and thus slightly above the top of
the $d_{xy}$ band at $\bar{\Gamma}.$ Hence, there are two \={M}-centered
hole pockets. As seen from the color coding in Fig. \ref{fig:models}, the
inner pocket has $d_{xz}$ character towards \={Y} $\left( 0,\pi \right) $
and $d_{yz}$ character towards \={X} $\left( \pi ,0\right) ,$ while the
opposite is true for the outer pocket. Towards $\bar{\Gamma},$ the more
dispersing band forming the inner pocket has longitudinal, i.e. $d_{Xz}$
character ($X$ is the $\bar{\Gamma}$\={M} direction)
, while the less dispersive band forming the outer pocket
has transversal, i.e. $d_{Yz}$ character.

As we move away from its top at $\bar{\Gamma},$ the pure $d_{xy}$ band
decreases by about 0.5 eV to minima at \={X} $\left( \pi ,0\right) $ and \={Y%
} $\left( 0,\pi \right) ,$ and towards \={M} it decreases even more, but
then hybridizes strongly and looses its $d_{xy}$ character. Starting now
from the minima at \={X} and \={Y}, and going towards \={M}, hybridization
with As $p_{z}$ which increases linearly with the distance from \={X} or \={Y%
} makes the $d_{xy}/p_{z}$ antibonding band disperse strongly \emph{up}%
wards, cross the Fermi level, and near \={M} reach a maximum which in
LaOFeAs is more than 1 eV above the Fermi level and in SmOFeAs is $\sim $0.3
eV lower.

Similarly, moving away from the degenerate top at \={M}, the nearly pure $%
d_{xz}$ band decreases slowly towards \={X} where it reaches a minimum
merely 0.1 eV below the Fermi level. Going from there towards $\bar{\Gamma}$%
, hybridization with As $p_{y}$ makes the $d_{xz}/p_{y}$ antibonding band
disperse strongly upwards, cross the Fermi level, and reach its maximum at $%
\bar{\Gamma},$ 2 eV higher. Together with the above-mentioned $d_{xy}/p_{z}$
band, the $d_{xz}/p_{y}$ band forms an \={X}-centered electron pocket whose
shape is like a super-ellipse pointing towards \={M}. The two bands have
different minima at \={X} and cross along \={X}\={M}, because along this
line they are not allowed to hybridize with each other. In the lower plots
of Fig. \ref{fig:models} we thus see that the \={X}-centered electron pocket
has predominant $d_{xz}$ character on the long sides, and $d_{xy}$ character
along the short sides. Analogously for the \={Y}-centered electron pocket
which has $d_{yz}/p_{x}$ character along the long sides and $d_{xy}/p_{z}$
character along the short sides.

The essential difference between the band structures of Sm- and LaOFeAs is
that due to the increased As height in Sm, the hopping integral, $t_{xy,z},$
between As $p_{z}$ and Fe $d_{xy}$ is decreased, whereby the antibonding $%
p_{z}$ bymixing to the $d_{xy}$ band from \={X} or \={Y} and towards \={M}
is diminished, and herewith the slope of that band. This makes the electron
pockets in Sm- longer --and the Fermi velocity on the short sides lower--
than in LaOFeAs. This is illustrated in Fig. \ref{fig:models} and was
discussed in Sect. 4.1 of Ref.$\,$\onlinecite{AndersenBoeri2011}.

Andersen and Boeri\cite{AndersenBoeri2011} noticed that the antibonding $%
d_{xy}/p_{z}$ band near \={M} causes the dispersion of the inner $%
d_{xz},d_{yz}$ band to be linear over an energy range which is the larger,
the closer the $d_{xy}/p_{z}$ level is to the degenerate $d_{xz},d_{yz}$
level at \={M}. The reason is, that the hybridization between $d_{xy}/p_{z}$
and $d_{xz},d_{yz}$ increases linearly with the distance from \={M}, so that
if the $d_{xy}/p_{z}$ level were degenerate with the $d_{xz},d_{yz}$ level,
then the $d_{xy}/p_{z}$ band would form a Dirac cone with one of the $%
d_{xz},d_{yz}$ bands, which then becomes the inner band. Both for La and
SmOFeAs, this linear dispersion extends around the Fermi level. In the
present functional RG calculations, no effect of this strong and linear
dispersion was found.

Lacking an ab-initio 8-orbital one-electron Hamiltonian for SmOFeAs, we used
the one for LaOFeAs and reduced $t_{xy,z}$ from 520 to 325 meV to fit the
DFT-GGA-LAPW bandstructure of SmOFeAs. The latter is shown in Fig. 9 of Ref.$%
\,$\onlinecite{AndersenBoeri2011} and the result of the model is shown on
the upper right-hand side of Fig. \ref{fig:models}.

Concerning the Coulomb correlations, we included the density-density terms
for the onsite interactions, $U_{dd}$ and $U_{pp},$ between the different $p$
or $d$-orbitals on a given As or Fe-site, as well as the repulsion, $U_{pd},$
between the $d$- and $p$-orbitals on neighboring Fe- and As-sites. In
addition, onsite exchange and onsite pair-hopping interactions of the $d$%
-orbitals were included. The numerical values were taken from cRPA
(constrained random phase approximation) studies by Miyake et al.\cite%
{Miyake2010}. 
They are displayed in Table \ref{tab:int_values}.
Rigorously viewed, building the model Hamiltonian from results
of two different ab-initio calculations could lead to inconsistencies, but
we shall argue in Sect.$\,$\ref{sec:La1111} that this does not affect our
qualitative results.

By the transformation to the $\left( \mathbf{k},b\right) $-\emph{band}representation
we get the expression
\begin{align}
H_{I,\Lambda _{0}}=& \frac{1}{N}\sum_{\mathbf{k_{1}},\mathbf{k_{2}},\mathbf{%
k_{3}}}\sum_{{b_{1},\ldots ,b_{4}}\atop{s,s^{\prime }}}V_{\Lambda _{0}}(%
\mathbf{k_{1}},b_{1};\mathbf{k_{2}},b_{2};\mathbf{k_{3}},b_{3};b_{4})\cdot 
\notag \\
& \cdot a_{\mathbf{k_{3}},b_{3},s}^{\dagger }a_{\mathbf{k_{1}}+\mathbf{k_{2}}%
-\mathbf{k_{3}},b_{4},s^{\prime }}^{\dagger }a_{\mathbf{k_{2}}%
,b_{2},s^{\prime }}a_{\mathbf{k_{1}},b_{1},s}
\end{align}%
where the index $\Lambda _{0}$ is used to point out that the
electron-electron couplings are the bare ones with respect to our basis.
Since we are mainly interested in the consequences of the band structure
differences, we first use the LaOFeAs interaction parameters also in the
Sm-case, and do some additional checks later on.

\begin{table}
    \centering
    \begin{tabular}{l | c c c c c}
        $U$       & $x^2-y^2$ & $Yz$   & $3z^2-1$ & $Xz$   & $xy$ \\ \hline
        $x^2-y^2$ & $4.66$ &    $3.09$ & $2.99$ &   $3.09$ & $3.57$ \\
        $Yz$      & $3.09$ &    $4.08$ & $3.31$ &   $2.90$ & $2.91$ \\
        $3z^2-1$  & $2.99$ &    $3.31$ & $4.33$ &   $3.31$ & $2.81$ \\
        $Xz$      & $3.09$ &    $2.90$ & $3.31$ &   $4.08$ & $2.91$ \\
        $xy$      & $3.57$ &    $2.91$ & $2.81$ &   $2.91$ & $3.98$ \\ \hline\hline
        $J$       & $x^2-y^2$ & $Yz$   & $3z^2-1$ & $Xz$   & $xy$ \\ \hline
        $x^2-y^2$ &        &    $0.63$ & $0.74$ &   $0.63$ & $0.37$ \\
        $Yz$      & $0.63$ &           & $0.45$ &   $0.56$ & $0.59$ \\
        $3z^2-1$  & $0.74$ &    $0.45$ &        &   $0.45$ & $0.67$ \\
        $Xz$      & $0.63$ &    $0.56$ & $0.45$ &          & $0.59$ \\
        $xy$      & $0.37$ &    $0.59$ & $0.67$ &   $0.59$ &        \\
    \end{tabular}
    \caption{The table shows parameters for onsite interactions of the Fe $d$ orbitals in eV, from Ref. \onlinecite{Miyake2010}. 
        Density-density terms are given in the upper half, while the lower half shows the
        values for exchange interactions. Concerning the onsite density-density interaction 
        of the As $p$ orbitals and nearest neighbor $p$-$d$ repulsion we use the values 
        $U_{pp}=2.6$ eV and $U_{pd}=1.2$ eV.}
    \label{tab:int_values}
\end{table}

\section{Functional RG Method}

\label{sec:method} The functional RG approach we use here is basically the
same as the one used previously for one-band Hubbard models\cite%
{ZanchiSchulz2000,PhysRevB.61.7364,honerkamp2001} and five-band models for the iron arsenides\cite%
{FaWang2009,ThomaleMay2011,ThomaleSep2011}. The underlying formalism and the
most important previous applications are discussed in some detail in a
recent review by Metzner et al.\cite{metznerRMP} as well as, with a
particular emphasis on iron-based superconductors, in a review by
Platt et al.\cite{PlattReview}. Here we only sketch the
main points.

The goal of the fRG approach is to derive an effective low-energy
interaction that couples the fermionic degrees of freedom near the Fermi
surfaces, say below an energy scale $\Lambda$. This effective interaction
should include higher-order corrections due to the excitations at higher
energies. These can be included by integrating out the higher-energy modes
in the functional integral. In a scheme that is perturbative in the
interactions (which should be a reasonable choice for the iron arsenide
superconductors) this requires the summation of an infinite number of
diagrams. One strength of the RG approach is that it replaces the summation
of infinitely many diagrams with all higher-energy modes included by a
differential equation whose right hand side contains a handful of
second-order diagrams with only modes above an energy scale $\Lambda$ contributing.
More precisely, one computes the change of the effective interaction $%
V_\Lambda (\mathbf{k_1},b_1;\mathbf{k_2},b_2;\mathbf{k_3},b_3;b_4)$ for
incoming electrons specified by wave vectors $\mathbf{k_1}_{/\mathbf{2}}$
and band indices $b_{1/2}$ and outgoing electrons $\mathbf{k_3}_{/\mathbf{4}%
} $, $b_{3/4}$ (with $\mathbf{k_4}$ being fixed by momentum conservation)
with decreasing bandwidth $\Lambda$ of the low-energy effective theory, i.e.
with successively including the modes with band energies above $\Lambda$. In
this notation, spin-rotational invariance is understood and the spin
components along the respective quantization axis of particles 1 and 3 and
those of 2 and 4 are pairwise identical. The RG equation is then given by 
\begin{equation}  \label{equ:flow_equ}
\frac{d}{d\Lambda } V_\Lambda (\mathbf{k_1},b_1;\mathbf{k_2},b_2;\mathbf{k_3}%
,b_3;b_4) = \sum_{\mathbf{k},\omega, b, s} V_\Lambda \otimes L_\Lambda
\otimes V_\Lambda \, .
\end{equation}
In this functional differential equation the summation is over internal
quantum numbers (loop variables) $\mathbf{k},\omega,b,s$ for wave vector,
Matsubara frequency, band and spin index, and all five one-loop diagrams $%
L_\Lambda$, which are one particle-particle bubble and four different
particle-hole diagrams built from two free propagators. The symbols $\otimes$
indicate the convolution of these loop variables with the external variables 
$\mathbf{k_1},b_1,\mathbf{k_2},b_2, \mathbf{k_3},b_3,b_4$ in the arguments
of the functions on this right hand side. As we use a momentum-shell cutoff
that excludes the modes below scale $\Lambda$, one of the two propagators is
at the cutoff energy, i.e. has its absolute value of the band energy, $|\epsilon|$, equal
to $\Lambda$, while in the other propagator, the band energy $%
\epsilon^{\prime}$ fulfills $|\epsilon^{\prime}| \ge \Lambda$. More details
can be found in Refs.~\onlinecite{metznerRMP,PlattReview}.

Let us also comment on the approximations that go into this fRG scheme.
First of all, in the exact hierarchy of flow equations for the one-particle
irreducible (1PI) vertex functions, on the right hand side there would be
another term representing the impact of the 1PI six-point vertex on the
effective interaction (also called 1PI four-point vertex). The 1PI six-point
vertex is zero in the bare action but would be generated during the flow. In
our truncation we simply drop this term. The impact of the six-point vertex
has only been explored recently, and it was found that in simple models it
can have at best quantitative but no qualitative effects\cite{Maier2012}. As the usual Fermi
surface instabilities can be understood without these higher order vertices,
we think that it is a safe approximation to employ this truncation here. 
The next and possibly more severe approximation is to drop the self-energy effects on the right hand side
of the fRG equation for $V_\Lambda$. While the self-energy is definitely
very important close to the instability, the accumulated body of experiences
for the one-band Hubbard model shows that the leading flows are not altered
qualitatively if one includes (parts of) the self-energy feedback. 
On the other hand, for multi-band systems with multiple Fermi surfaces, the situation could be more difficult, as e.g. small Fermi pockets could be closed already by weaker self-energy effects. The issue of self-energy effects in many-body approaches built on DFT band structures would  in addition be overshadowed by the question of a possible double-counting of interaction effects that have already been included in the DFT. This problem must be dealt with in further research.
Finally, we also neglect the
frequency dependence of the effective interactions. This is again absent in
the bare action, unless one would implement the frequency dependence of cRPA interaction parameters\cite{aryasetiawan}.
During the fRG flow, the one-loop diagrams would create a frequency dependence on the interaction on three wavevectors.
However, as we are interested in static instabilities at low $T$, we
can focus on the flow of the vertices at the smallest fermionic
Matsubara frequencies, which go to $0$ as $T \to 0$. Hence in
$V_{\Lambda}$ on the left hand side of the flow equation \ref{equ:flow_equ}, all the
fermionic Matsubara frequency are treated as being zero. On the right
hand side of \ref{equ:flow_equ}, the Matsubara sums of the loop diagram still contain
two electron propagators whose frequency dependencies are kept. However, since
for the vertices we use the zero-frequency values, we can do these sums
analytically. Recent fRG studies of the one-band Hubbard\cite{uebelacker,giering} model
take the full frequency dependence of the vertices into account,
without major qualitative changes for the context considered here.

\begin{figure}[tbp]
\includegraphics[width=0.3\textwidth]{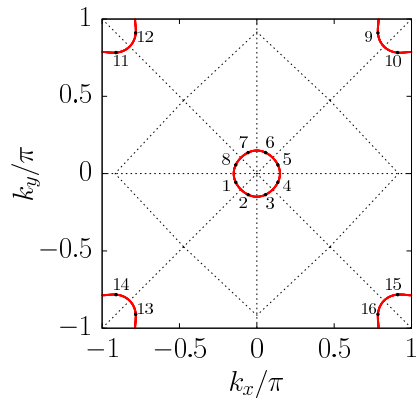}
\caption{The BZ with two Fermi surface pockets from the same band is divided
into $2\cdot m$ patches, with $m=8$ in this example. Every patch contains
one patch point on the corresponding Fermi surface.}
\label{fig:ShowPatch}
\end{figure}

\begin{figure}[tbp]
\includegraphics[width=0.23\textwidth]{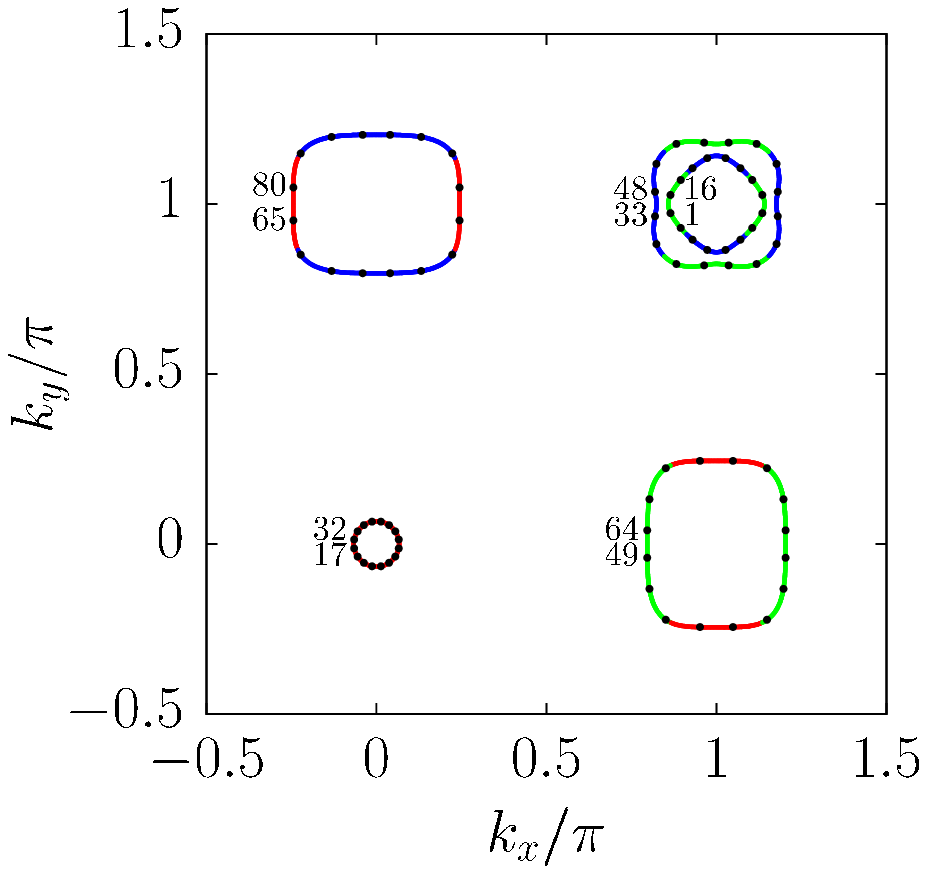} %
\includegraphics[width=0.23\textwidth]{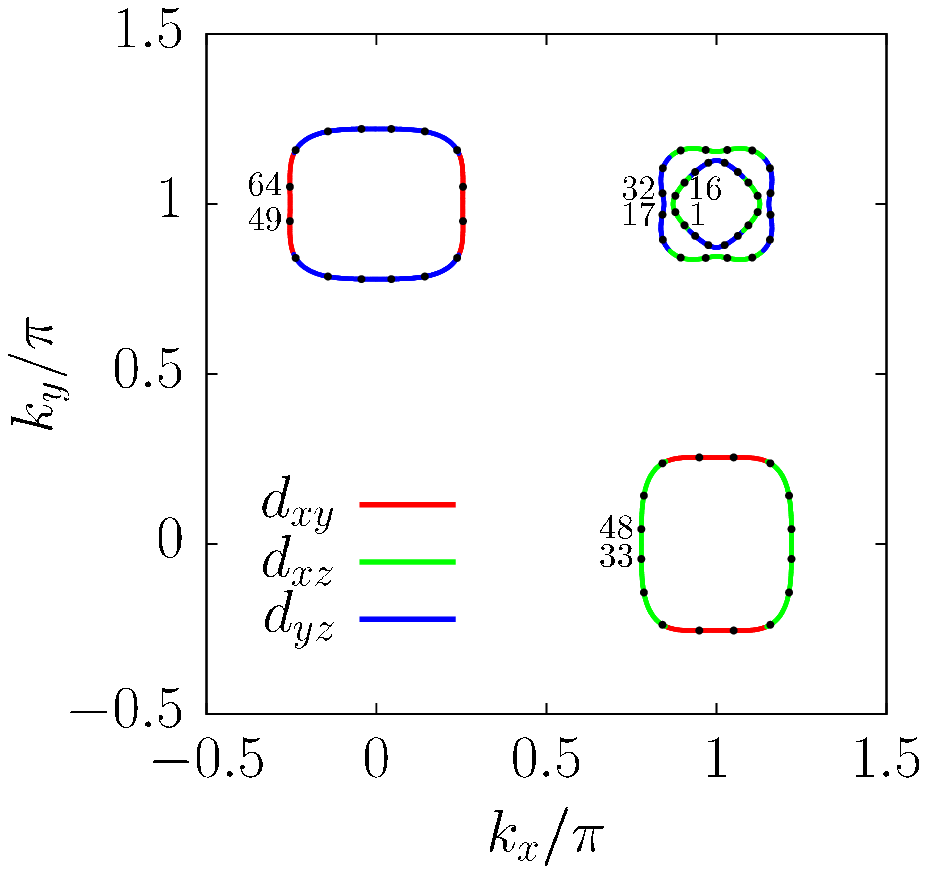}
\caption{Fermi surface with patch point configuration and numbering for
La-1111 with electron doping $x=0.10$ (left) and $x=0.15$ (right).}
\label{fig:PatchPoints}
\end{figure}

Besides these simplifications of the fRG equation, we also need an
appropriate discretization of the momentum space to be able to integrate
this differential equation numerically. For this purpose we use a Fermi
surface patching, which was described for the first time by Zanchi and Schulz%
\cite{ZanchiSchulz2000}. Concerning a band that forms at least one Fermi
surface, the Brillouin zone is thereby divided into several patches, the
number of which is equal to $m$ times the number of Fermi-surface sheets of
this band. The arrangement of the patches and corresponding patch points is
illustrated in Fig. \ref{fig:ShowPatch}. Now we approximate the interaction
to be constant within one patch and use the value at the Fermi surface patch
point as a patch value. This approximation can be understood as the mapping 
\begin{equation}
V_{\Lambda }(\mathbf{k_{1}},b_{1};\mathbf{k_{2}},b_{2};\mathbf{k_{3}}%
,b_{3};b_{4})\rightarrow V_{\Lambda }(i_{1},i_{2},i_{3},b_{4})\mathnormal{,}
\end{equation}%
where $i_{j}$ is the patch number corresponding to wave vector $\mathbf{k_{j}%
}$ and band $b_{j}$. A higher value of $m$ means higher resolution, but also
causes higher computational effort. Hence we used $m=16$ as a reasonable
compromise for our calculations. The arrangement of the patch points for
La-1111 is shown in Fig. \ref{fig:PatchPoints} in the case of a moderate
doped system with five pockets and of a higher doped system where the $%
d_{xy} $ hole pocket at $\bar{\Gamma}$ is gone. For Sm-1111 this arrangement
is very similar to the one shown here. Within this treatment we do not take
bands into account that do not cross the Fermi level. This helps to reduce
the numerical effort. If such neglected bands are well separated from the
Fermi level, this approximation may be very good. If the separation in
energy gets small, it can lead to quantitative inaccuracies, especially when
there is a band that runs close to the Fermi level without crossing it. The
effects of those inaccuracies become evident at some places in our results, as discussed below.
For example, the doping-driven transition from a five-pocket to a four-pocket system comes out rather abrupt, while it can be expected to be much smoother in a discretization scheme with a higher resolution of the Brillouin zone and with additional patch points away from the Fermi surface.

Our implementation of the fRG procedure given above is based on one that has
been used for previous publications by some of the authors (see e.g. Ref.\onlinecite{ThomaleMay2011}). Besides an
extension of the code to eight-orbital/-band systems, we
have also implemented improvements concerning some details of the
calculation. One detail that should be mentioned here is that the coupling
is stored in two different representations during the flow. In addition to
the band-dependent effective interaction, $V_\Lambda(i_1,i_2,i_3,b_4)$, we
can also transform to the orbital-dependent interaction $V_%
\Lambda(i_1,i_2,i_3,o_4)$ with an orbital index $o_4$ instead of a band
index $b_4$ on the fourth leg. 

Taken together, the fRG in the current setup that is used here contains a
number of approximations which could, in bad cases, certainly lead to severe
quantitative misestimations of critical scales. On the other hand doing less
approximations for multi-band systems will remain difficult for a while, and
therefore it is important to learn to what extent the so-obtained
information complies with experiments and other methods. In this paper we
study whether an experimental $T_{c}$-trend is rendered correctly by the fRG
in this form. The result is promising, indicating the possibility
that the quantitative artifacts of the approximations are similar for the
different systems studied. This provides confidence that the fRG description
makes sense and can be extended to other questions.

Before the results for La-1111 and Sm-1111 are shown, we will briefly
explain how the renormalized couplings at lower scales can be analyzed. When some values of
the effective interaction begin to diverge near a critical scale $\Lambda
_{c}$, we have to stop the renormalization and perform a mean-field
analysis for the modes below $\Lambda_c$. This renormalized mean-field analysis has also been done by Reiss
et al.\cite{ReissRoheMetzner2007} in a more complex way and is explained in
their paper(see also Refs. \onlinecite{platt_sid,PlattReview,WangEberlein} for further examples and developments). In a simple description we can assume that in the case of a
leading Cooper instability we can focus on the effective Hamiltonian restricted to the modes below $\Lambda_c$
\begin{equation}
H=\sum_{\mathbf{k},s}\epsilon _{\mathbf{k}}a_{\mathbf{k},s}^{\dagger }a_{%
\mathbf{k},s}+\frac{1}{N}\sum_{\mathbf{k},\mathbf{k^{\prime }}}V_{\mathbf{k},%
\mathbf{k^{\prime }}}^{\Lambda _{c}}a_{\mathbf{k},\uparrow }^{\dagger }a_{-%
\mathbf{k},\downarrow }^{\dagger }a_{-\mathbf{k^{\prime }},\downarrow }a_{%
\mathbf{k^{\prime }},\uparrow }
\end{equation}%
with $V_{\mathbf{k},\mathbf{k^{\prime }}}^{\Lambda _{c}}=[V_{\Lambda _{c}}(%
\mathbf{k^{\prime }},-\mathbf{k^{\prime }},\mathbf{k})+V_{\Lambda _{c}}(-%
\mathbf{k^{\prime }},\mathbf{k^{\prime }},\mathbf{k})]/2$, with the wavevectors in the corresponding bands. 
By introducing the BCS mean-field $\Delta _{\mathbf{k}}=\frac{1}{N}%
\sum_{\mathbf{k^{\prime}}}V_{\mathbf{k},\mathbf{k^{\prime }}}^{\Lambda
_{c}}\langle a_{-\mathbf{k^{\prime }},\downarrow }a_{\mathbf{k^{\prime }}%
,\uparrow }\rangle $ and neglecting higher order fluctuations the
Hamiltonian becomes quadratic whereby it can be diagonalized easily. The
value of the gap function has to be calculated self-consistently from the
gap equation, which can be simplified to the linearized gap equation 
\begin{equation}
\Delta _{\mathbf{k}}=-\frac{1}{N}\sum_{\mathbf{k}^{\prime }}V_{\mathbf{k},\mathbf{%
k^{\prime }}}^{\Lambda _{c}}\frac{\Delta _{\mathbf{k^{\prime}}}}{2\epsilon _{%
\mathbf{k^{\prime}}}}\tanh \left( \frac{\epsilon _{\mathbf{k^{\prime}}}}{%
2T_{c}}\right)
\end{equation}%
in the limit $T\rightarrow T_{c}$. Now the patch discretization with patches $i$ within which the gap is held constant can be used
to simplify this to an eigenvalue problem 
\begin{equation}
\eta \Delta _{i}=\sum_{j}V_{i,j}^{\Lambda _{c}}\Delta _{j} \label{equ:eigenvalueP}
\end{equation}%
where we have assumed that the integrations within each patch, $-\frac{1}{N}\sum_{\mathbf{k}^{\prime } \in i} \frac{\tanh \left( \frac{\epsilon _{\mathbf{k^{\prime}}}}{%
    2T_{c}}\right)}{2\epsilon _{\mathbf{k^{\prime}}}}\approx - \frac{\rho_0}{n} \ln \frac{1.13 \Lambda_c}{T_c}=:\eta^{-1} $, do not depend strongly on the patch index $i$. $\rho_0$ is the density of states at the Fermi level, and $n$ is the number of patches.
The solution requires that $\Delta_i$ is an eigenvector of $V_{i,j}^{\Lambda _{c}}$, with eigenvalue $\lambda = \eta$.
The patch dependence of the eigenvector controls the gap form factor, i.e.  the angular dependence of the gap.
The result for the eigenvalue can be resolved
for the critical temperature, yielding 
\begin{equation}
T_{c}=1.13 \Lambda _{c} \, e^{\frac{n}{\lambda \rho _{0}}}\mathnormal{,%
}
\end{equation}%
which is maximal for the lowest (or most negative) eigenvalue. Thus the
solution with the lowest eigenvalue characterizes the ground state
properties of the system. Note that in this context, the eigenvalue $\lambda$ scales linearly with  the patch number. 
Other instabilities, like the spin-density-wave
(SDW), can be analyzed analogously. By comparing the eigenvalues the leading
instability can be identified. Moreover, an important implication of the last equation is that the
scale $\Lambda _{c}$ is a measure for the critical temperature so that we
can use it to compare the transition temperatures of different systems in a
simple way.


\section{Results for LaOFeAs}

\label{sec:La1111} 
\begin{figure}[tbp]
\includegraphics[width=0.46\textwidth]{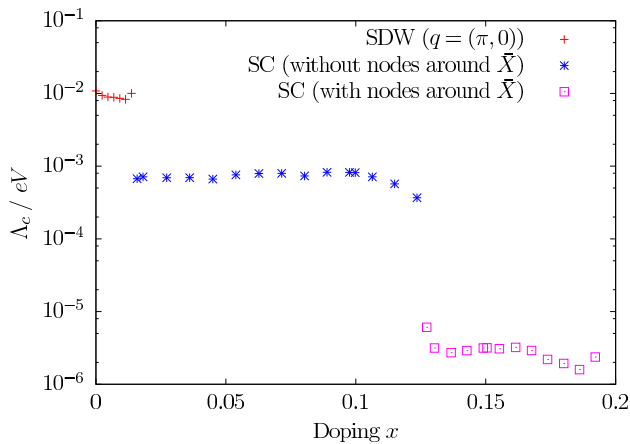}
\caption{FRG phase diagram for La-1111. The colors or symbols indicate the
leading instability at a given electron doping.}
\label{fig:LaPhaseDiag}
\end{figure}

First, let us describe the fRG results for the La-1111
parameters. We find flows to strong coupling over a wide parameter range.
The leading couplings are either in the SDW channel with wave vector
transfer $\sim (\pi ,0)$ or in the pairing channel for zero total incoming
momentum. In Fig. \ref{fig:LaPhaseDiag} we plot the critical scale versus
doping parameter $x$, together with the information what the leading channel
is. For small doping, the SDW instability prevails, while beyond a critical
doping the superconducting instability is stronger. The clear separation of
these two phases is emphasized by the step in the critical scale which has a
similar magnitude as the one found in experiments by Luetkens et al.\cite%
{Luetkens2009}. There is another step in $\Lambda _{c}$ inside the
superconducting regime exactly where the hole pocket at $\bar{\Gamma}$
vanishes. 
It has not been seen in experiments and we have proved that it
originates in our implementation at least partially which ignores energy
bands not crossing the Fermi level. For this purpose we took into
account points for the $\bar \Gamma$ band also in the overdoped regime
and observed a smoother behavior at the transition from five to four
pockets.
However, it is clearly physically sound that the loss of a Fermi pocket reduces the pairing scale. 

\begin{figure}[tbp]
\includegraphics[width=0.22\textwidth]{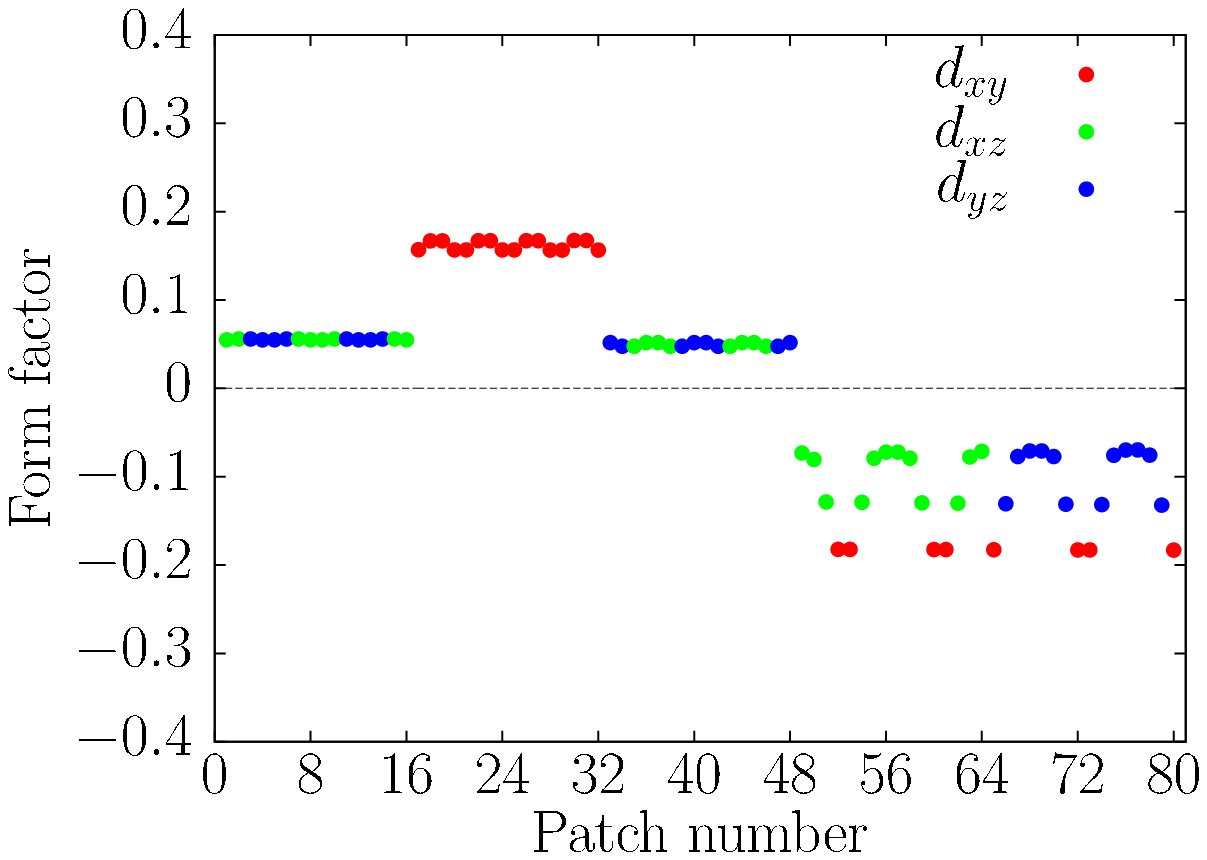} %
\includegraphics[width=0.22\textwidth]{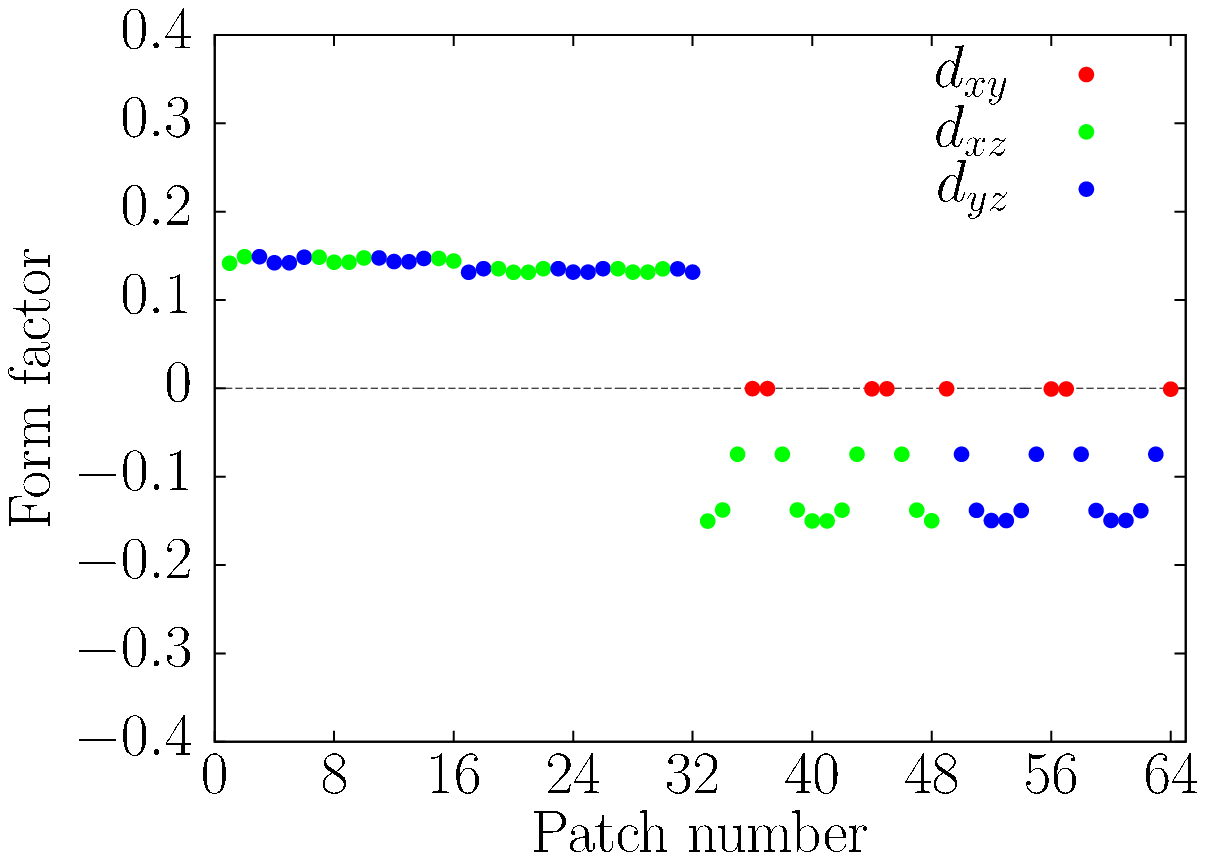}
\caption{FRG form factors of the superconducting gap for La-1111 at electron
dopings $x=0.10$ (left) and $x=0.15$ (right) calculated from the linearized
gap equation.}
\label{fig:LaGaps}
\end{figure}

In the superconducting regime we have calculated gap form factors for
various dopings by solving the eigenvalue problem (\ref{equ:eigenvalueP}).
This leads to the normal decomposition:%
\begin{equation}
V_{i,j}^{\Lambda _{c}}=\sum_{l}\lambda _{l}f_{i}^{\lambda
_{l}}f_{j}^{\lambda _{l}}\approx \lambda _{min}f_{i}^{\lambda
_{min}}f_{j}^{\lambda _{min}},
\end{equation}%
in terms of coupling strengths (eigenvalues), $\lambda ,$ and form factors
(eigenvectors) $\mathbf{f^{\lambda }}$.  The last approximation $\approx$ in the equation above keeps only
the term from the leading (most negative) eigenvalue. The components $f_i^{\lambda_{min}}$ of the
corresponding form factor, $\mathbf{f^{\lambda _{\min }},}$ are shown in
Fig. \ref{fig:LaGaps} and the numbering of the  points $i$ running around the Fermi surfaces in Fig. %
\ref{fig:PatchPoints}.
Fig. \ref{fig:LaGaps} shows two of these form factors, one for a five-pocket
and one for a four-pocket system, where the color indicates the dominant
orbital at each patch point on the Fermi surface. Both exhibit a $s^{\pm }$%
-symmetry and a strong gap anisotropy around the electron pockets, but gap
nodes only occur in the four pocket case. All systems studied in the five-pocket
regime show a similar form factor to the one shown in the left plot of Fig. \ref{fig:LaGaps}, 
and all systems in the four-pocket regime agree qualitatively with the data shown in the right plot. 
This means that the $\bar{\Gamma}$ pocket is necessary for a
diverging $d_{xy}$ interaction, which dominates in the five pocket-systems
and leads to high critical scales but is negligible at four pockets. 
However, as in the case of the critical scale the sudden form factor
change between five and four pocket regime is smoothed out by including
the $\bar \Gamma$ band in the calculations for overdoped systems.
When we
compare this to previous five-band results\cite{ThomaleMay2011}, a different
anisotropy around the electron pockets is striking in the five pocket model
for La-1111. In these older findings the dominant interactions had $d_{xz}$
and $d_{yz}$ character, which resulted in an inverted anisotropy compared to
the one shown here. In contrast, the form factors of the four-pocket systems
are very similar. Since the interactions that involve $p$-orbitals are not
taken into account explicitly in five-band studies, we analyze the effect of
the initial interactions of this kind on our system. It becomes apparent
that the dominance of the $d_{xy}$-orbital in the form factor for the
five-pocket systems is reduced when one neglects initial $p$-$d$
interactions. This behavior suggests that the difference in the gap
anisotropy between the five- and eight-band results is systematic due to
downfolding on distinct basis sets. 
The orbital composition of the superconducting pairing can hardly be
measured, but the differences in orbital composition of the pairing changes
the gap anisotropy quantitatively, which is a measurable effect. Note that
in case of a dominant $d_{xy}$ pocket, the gap minima on the electron
pockets appear on the long sides, while in the five-band studies, they are
located on the short sides\cite{ThomaleMay2011}. We have checked that is
difference not simply matter of slightly different parameters for the two
models. By reducing the relative strength of the $U_{xy,xy}$-interaction
parameter compared to the other $U$s, we can tune the five-pocket case in the eight-band model to give
gap structures analogous to the five-band results, but we have to change these
interaction parameters substantially for this.

In order to get more insight into the structure of the renormalized couplings, we
transform Eq. (\ref{equ:eigenvalueP}) from band to orbital space and
diagonalize the matrix of pairing interactions between Fermi-surface
patches.
Now we can clearly see what was mentioned above: When the $d_{xy}$ hole
pocket is present, the pairing interactions between Fermi-surface points
with prevailing $d_{xy}$ character dominate and the gap minima on the
electron pockets appear on the long sides where the \={X} pocket has $d_{xz}$
and the \={Y} pocket $d_{yx}$ character. When the $d_{xy}$ hole pocket is
doped away and $d_{xy}$ character only remains on the short sides of the
electron pockets, the pairing interactions among $\mathbf{k}$-points with
prevailing $d_{xz}$ or $d_{yz}$ character dominate and the gap minima on the
electron pockets appear on the short sides where the character is $d_{xy}.$


Another interesting aspect is the sign structure of the form factors, which
supports the often stated $s^{\pm }$ symmetry. Note that in our formalism,
the signs of the coupling components in band representation are not
necessarily meaningful since they result from the choice of the arbitrary
signs in the transformation matrices from orbital to band picture.
Transformation to the orbital basis removes this ambiguity. With the
transformation chosen here, the sign structure of the gap turns out to be
the same, both in band picture as well as in the orbital basis. This
confirms the stated $s^{\pm }$ symmetry unambiguously.

Hence we conclude this comparison of the eight-band results with previous
five-band studies with the statement that the two models and their
fRG-implementations agree on the sign-changing $s$-wave pairing, but that
different gap anisotropies are possible. Different models have somewhat
different parameter regimes with respect to the leading orbital in the pairing and the location of the gap minima.

\begin{figure}[tbp]
\includegraphics[width=0.22\textwidth]{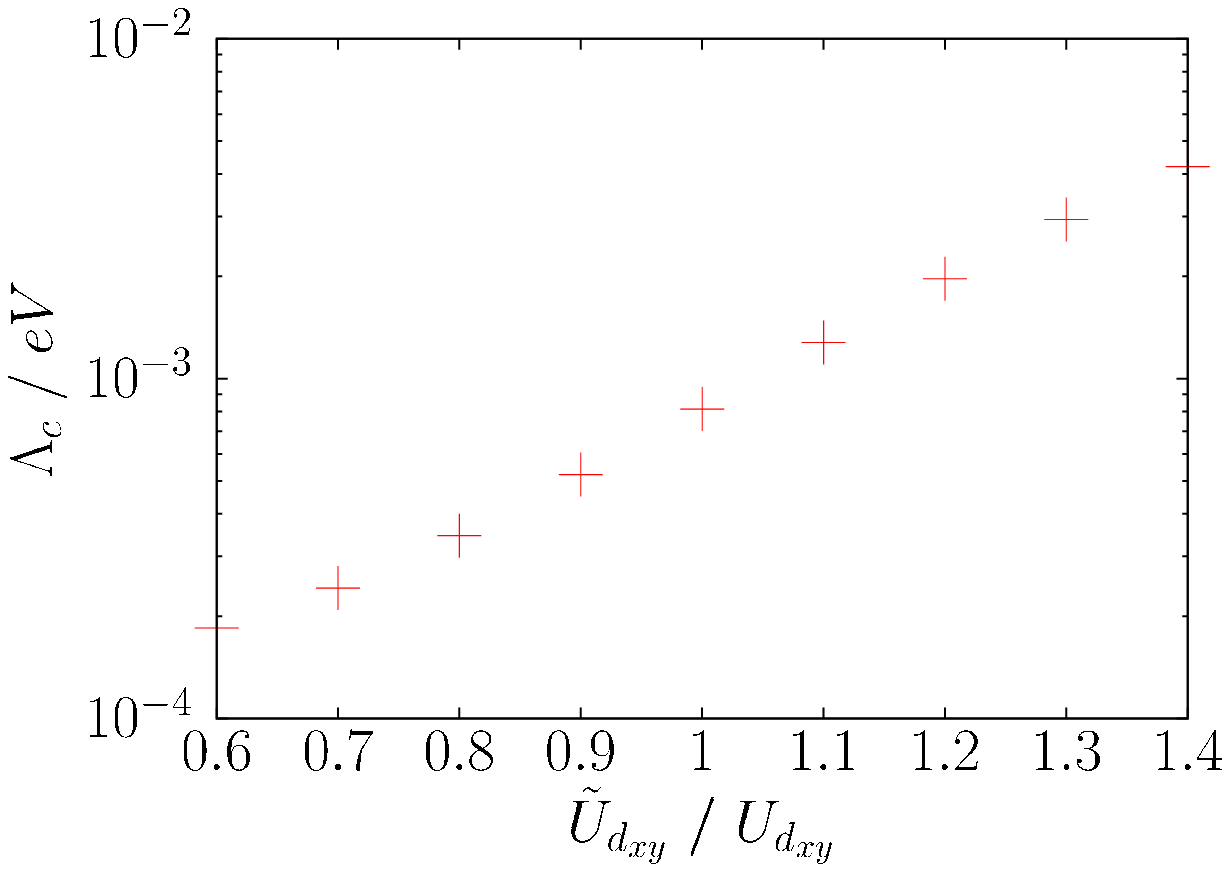} %
\includegraphics[width=0.22\textwidth]{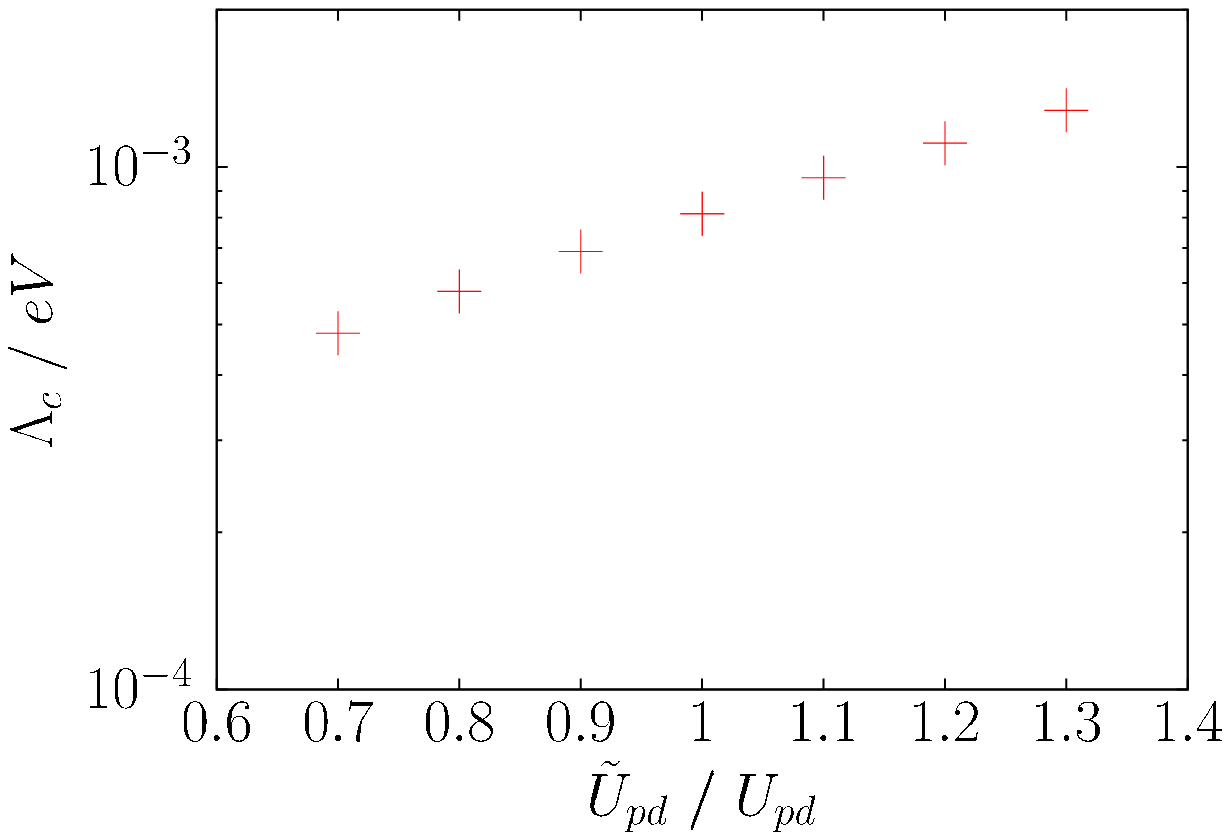}
\caption{The critical scales are plotted against varied initial interaction parameters $U_{\mu ,\mu ^{\prime }}\rightarrow \tilde{U}_{\mu,\mu ^{\prime }} = f\cdot U_{\mu,\mu ^{\prime }}$ with $\mu =\mu ^{\prime }=d_{xy}$ (left) or $\mu \in d$-orbitals and $\mu ^{\prime }\in p$-orbitals (right).}
\label{fig:vary_initUs}
\end{figure}

Next let us discuss the issue of combining results of two different
ab-initio calculations (one for the dispersion, one for the interaction
parameters) to build up our model. For this purpose we have changed the 
\emph{initial} interactions to analyze the stability of our results against
details of these values. More precisely we have added prefactors to some
interaction parameters $U_{\mu ,\mu ^{\prime }}\rightarrow f\cdot U_{\mu
,\mu ^{\prime }}$ in order to tune these values. We have varied a) the $pd$%
-interaction with $\mu \in d$-orbitals and $\mu ^{\prime }\in p$-orbitals,
and b) the onsite $dd$-interaction of the $\mu =\mu ^{\prime }=d_{xy}$%
-orbital. The latter one is known vary most between different iron pnictide
superconductors, according to cRPA-calculations\cite{Miyake2010} that
determine the effective interaction parameters for the model considered. In
both cases the resulting form factors have shown only a marginal dependence
on this prefactor. As shown in Fig. \ref{fig:vary_initUs} the critical 
scales are affected by changing these
prefactors, but their orders of magnitude not. In other words our,
qualitative results do not depend strongly on the details of the initial
interaction parameters. Thus, we think it is justified to use the above
given ab-initio model. Regarding the effect of the $U_{pd}$ interaction, we
have varied this value between $-50\%$ and $+150\%$ of the values listed in 
\onlinecite{Miyake2010} and found a monotonous trend. This shows that a positive $%
U_{pd}$ enhances the pairing scale, and mainly enforces the pairing within
the $d_{xy}$-dominated parts.

\section{Results for SmOFeAs}

\label{sec:Sm1111} 
\begin{figure}[tbp]
\includegraphics[width=0.46\textwidth]{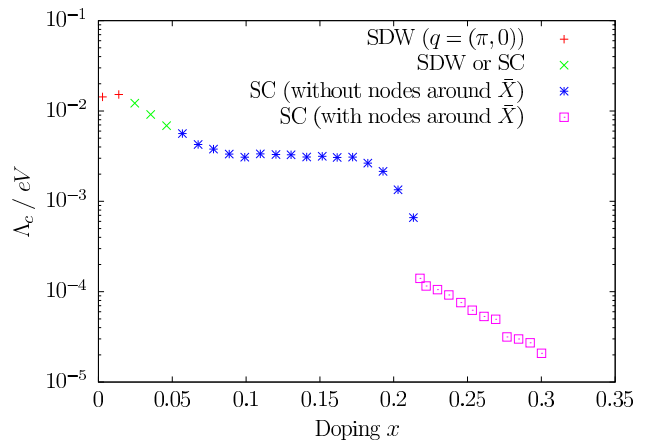}
\caption{FRG phase diagram for Sm-1111. The colors or symbols indicate the
leading instability at a given doping.}
\label{fig:SmPhaseDiag}
\end{figure}

In this part of the paper the fRG results for the Sm-1111
model are presented and compared with those for La-1111. Here again, we find
two competing instabilities in the investigated range of electron dopings.
In contrast to the phase diagram for La-1111, the data for Sm-1111 in Fig. %
\ref{fig:SmPhaseDiag} do not exhibit a step in the divergence scale
between the transition from the SDW to the SC phase. Furthermore, this
transition is not sharp for Sm-1111, but there is a region in which the
dominant instability cannot be determined unambiguously. Our findings are
consistent with the experimental results of Drew et al.\cite{Drew2009} which
also exhibit a transition between a magnetic and a superconducting phase
with continuous $T_{c}$ behavior. The absolute pairing scales in Sm-1111
from Fig. \ref{fig:SmPhaseDiag} are about three to four times as high as the
ones for La-1111 from Fig. \ref{fig:LaPhaseDiag} in the SC region, while
they are similar in the SDW region. This interesting parameter trend which
agrees with the experimental $T_{c}$s will be discussed further in the next
chapter.

\begin{figure}[tbp]
\includegraphics[width=0.22\textwidth]{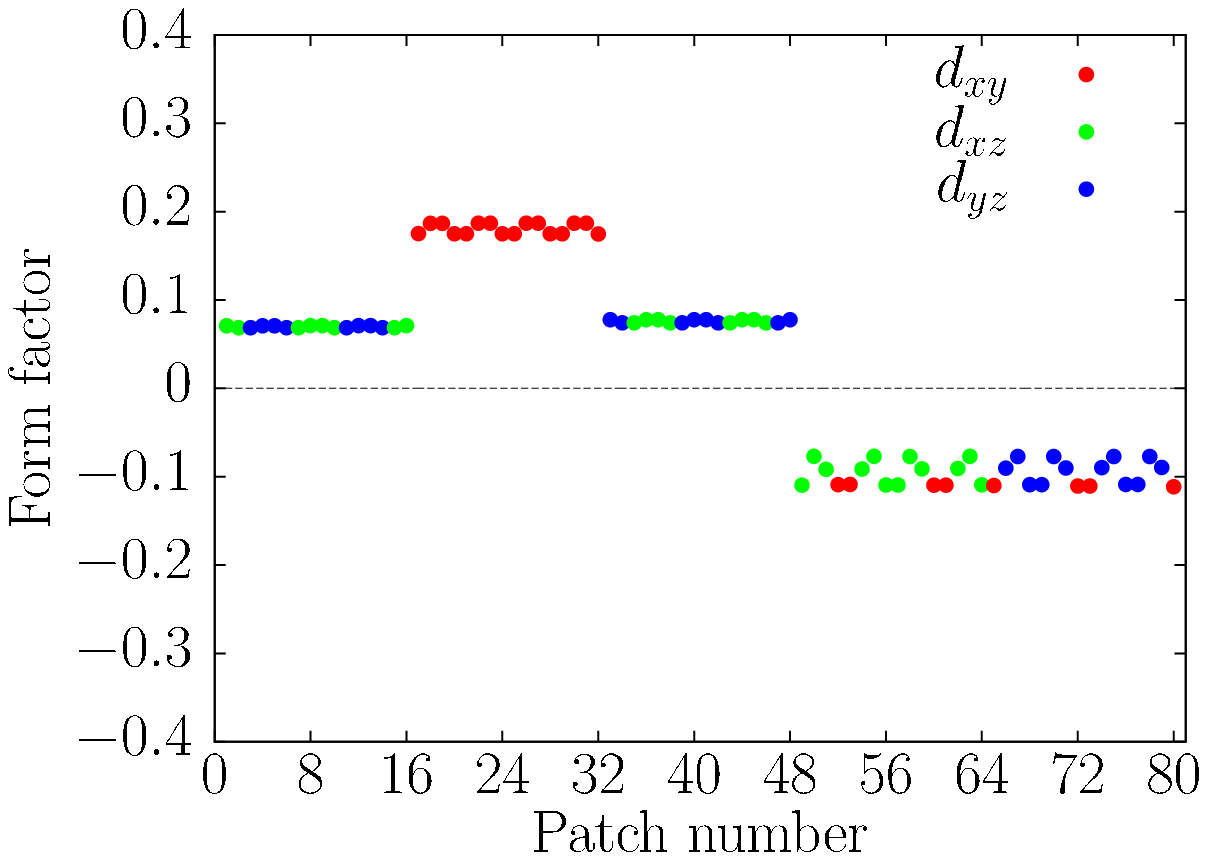} %
\includegraphics[width=0.22\textwidth]{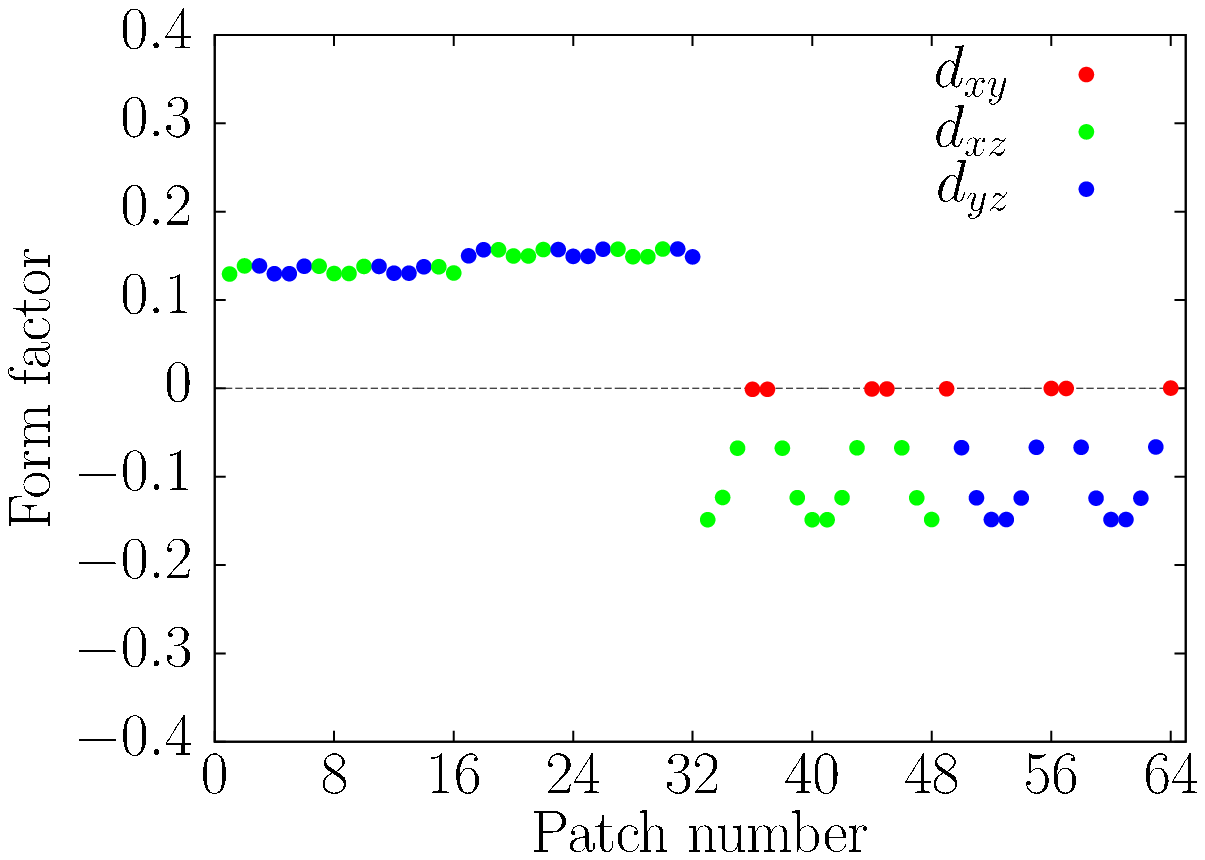}
\caption{FRG form factors of the superconducting gap for Sm-1111 at dopings $%
x=0.19$ (left) and $x=0.25$ (right) calculated from the linearized gap
equation.}
\label{fig:SmGaps}
\end{figure}

Like in La-1111, the superconducting form factors shown in Fig. \ref{fig:SmGaps} have $s^{\pm }$-symmetry
and depend mainly on whether the $d_{xy}$ pocket exists or not. In the
overdoped, four-pocket regime the form factors with nodes on the short, $%
d_{xy}$-dominated sides of the electron pockets, are nearly identical in the
two materials. In the optimally doped, five-pocket regime, the gap
anisotropy on the electron pockets is far less pronounced in
Sm-1111 and the position of the smallest gap has changed
on the long sides of the electron pockets. That
the gap on the electron pockets is far more isotropic in optimally doped
Sm-1111 than in La, should be experimentally observable.

\section{Material trend from LaFeAsO to SmFeAsO}

\label{sec:trend} We now concentrate on the calculated material-dependent
critical scales in the superconducting five-pocket regime. Our trend seems
to be consistent with the experimental observation that Sm-1111 has far
higher critical temperatures than La-1111. We shall distinguish between two
different effects of tuning the hopping integral $t_{xy,z}$ from its value
(520 meV) in La to the one (325 meV) in Sm, which could be responsible for
the above-mentioned material trend. The first effect stems from the change of the
eigenvectors of the Hamiltonian (\ref{BS}), i.e. of the orbital characters
on the Fermi surface.
These orbital characters enter the FRG flow via the initial conditions for the interactions 
expressed in the band picture. This means that the wavevector-dependence of the initial 
interaction depends on the tuning parameter $t_{xy,z}$.
The second place where the tuning parameter enters is the band energies, with direct impact on the
shape of the Fermi surface and the Fermi velocities.

\begin{figure}[tbp]
\includegraphics[width=0.46\textwidth]{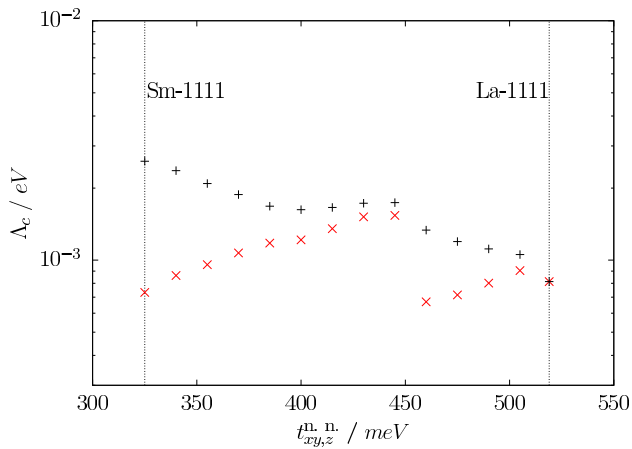}
\caption{Black crosses: Critical pairing scales of La-1111, Sm-1111 and fictitious intermediate systems versus
the band-structure tuning parameter, $t_{xy,z},$  at a
chemical potential of $\protect\mu =45\ \textnormal{meV}$, corresponding to $%
x=0.10$ in La-1111.  For comparison, with red crosses we show the critical
scales of fictitious systems whose dispersions are fixed to the La-1111 bands,
but whose composition of interactions changes according to the changing orbital character of the true $t_{xy,z},$-dependent bands.}
\label{fig:trendConstGeo}
\end{figure}

To analyze the effect of the orbital composition in the initial interaction, we have tuned the system
from La to Sm using the varying orbital-band transformation, but keeping the La
band structure, i.e. keeping the Fermi surfaces unchanged. Fig. \ref{fig:trendConstGeo} shows that the orbital
composition influences the critical scales, but cannot be responsible for
the higher scales in Sm because the trend is more or less opposite. The
discontinuity in the data in this case is presumably an artifact of the $N$%
-patch discretization (a near band crossing). For a better $\mathbf{k}$%
-discretization, this step should be smoothened out.

\begin{figure}[tbp]
\includegraphics[width=0.46\textwidth]{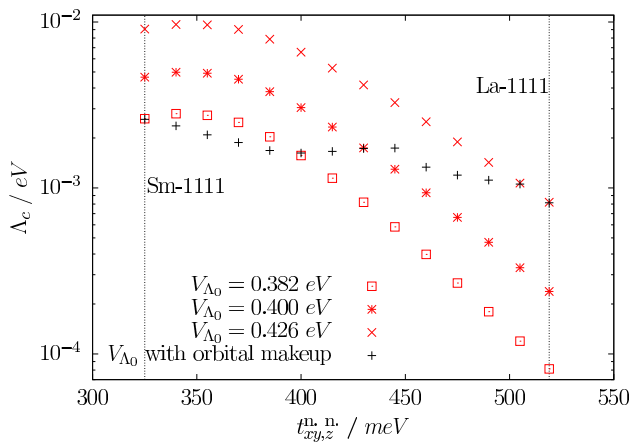}
\caption{Same as Fig. \protect\ref{fig:trendConstGeo}, except that the red
symbols are now for systems whose band structures vary with $t_{xy,z},$ but have
initial couplings, $V_{\Lambda _{0}},$ fixed to wavevector and bandindex independent values that are $382$, $400$ or $426$ meV.}
\label{fig:trendConstInitInt}
\end{figure}

In order to investigate the effect changing the band dispersions, we have
chosen 'averaged' initial interactions that do not depend on the band or on the wavevectors
 and then varied the band energies and the
Fermi surfaces via $t_{xy,z}$. The resulting critical scales are
shown in Fig. \ref{fig:trendConstInitInt} with three different average initial
coupling values. Here we see that the change of \emph{Fermi-surface shape}
and \emph{velocities} leads to increasing critical scales along the way from
La to Sm. Hence the material trend seems to be an effect of the band
dispersion.

The simplest attempt to interpret the effect of the dispersion may be to
consider it as a density-of-states effect. We have calculated the density of
states at the Fermi level in La-1111 as well as in Sm-1111 for the various
parameters under consideration. The result is that it is higher for Sm-1111 due
to the flattening of the $d_{xy}/p_{z}$ band forming the short sides of the
electron pockets. On first sight, this seems to be consistent with the
higher scales in the superconducting regime. However, this would also
suggest higher scales in the SDW phase, in contradiction to our fRG
findings, and also in contradiction to the experimental findings\cite{55K}.
So the density of states at the Fermi level appears to be too simple to give
a sufficient explanation. 

Next, we have computed the corresponding particle-hole diagrams at the
spin-density wave ordering vector and particle-particle diagrams at zero
total momentum. They sample the dispersion over a wider energy range than
the above-used density of states, which is only a measure at the Fermi
level. For Sm-1111, both bubbles at these specific wave-vectors become larger 
by a few percent in very similar way. So, naively, both pairing and SDW scale should go up.
In the fRG the larger SDW tendency due to a larger value of the particle-hole diagram could be compensated by a stronger screening of the repulsions in the likewise strengthened pairing channel. Yet, this does not help to understand in simple why the
pairing scale still manages to gain from the change in the band structure. 

Of greater significance might be that the
increased elongation of the Sm electron pockets caused by the flattening of
the $d_{xy}/p_{z}$ band, spreads out the $\left( \pi ,0\right) $ and $\left(
0,\pi \right) $ susceptibility peaks.
Indeed, by looking at the wavevector-dependence of the particle-hole bubble that determines the bare spin susceptibility, it becomes apparent that its peak not only becomes higher at the SDW ordering wavevector when going from La-1111 to Sm-1111, but it also becomes broader. We can use this susceptibility as pairing interaction in the BCS gap equation. Then the wavevector summation in the gap equation samples the whole neighborhood of the SDW peak, and one might hope that the broadening of the peak might help to understand the rise in the pairing scale. We have computed the pairing eigenvalues with the bare susceptibility as pairing interaction in the BCS gap equation. However, we found that the analysis of the parameter trend from La-1111 to Sm-1111 is severely disturbed by a number of competing pairing channels known from previous studies\cite{graser} that do not get removed unless we do a more sophisticated summation of many diagrams, as is ultimately done in the fRG.
 
Hence, we are led to the conclusion that it is rather difficult to use
single-channel or low-order considerations for the explanation of the
material trend from Sm-1111 to La-1111 seen in the fRG. A combined
treatment of all fluctuations channels to high orders such as in the fRG appears to be vital to capture this trend. 

\section{Discussion}

\label{sec:conclusions} We have performed extended fRG calculations for a
two-dimensional eight-band model for 1111 iron arsenides. The fRG goes
beyond RPA-type many-body approaches, and has until now only been applied to
five-band models for the iron superconductors. Our study shows that in the
eight-band model, the primary instabilities toward antiferromagnetic SDW and
sign-changing $s$-wave pairing are reproduced clearly. However, regarding
the question which orbitals contribute most to the pairing, we found some
model dependencies. In the eight-band $pd$ model with two electron and three
hole pockets, the $d_{xy}$ hole pocket plays a more important role for
pairing than in corresponding five-band $d$-only models. This also
qualitatively changes the gap anisotropy on the electron pockets. In the
eight-band model, the gap minima on these pockets appear on the long rather
than on the short sides. This difference should be experimentally
accessible. At present, we regard these differences as an indication of how
approximate many-body treatments of these complicated multi-band models currently are.
Future theoretical work must find ways to understand and reduce the model-dependence of the theoretical results.
On the positive side, we have seen that besides the $d_{xz/yz}$-dominated pairing found in five-band studies, also the second plausible scenario, the $d_{xy}$-dominated pairing is theoretically possible.  

We have used the hopping integral $t_{xy,z}$ between the Fe $d_{xy}$
and the As $p_{z}$ orbital to tune the eight-band model derived for La-1111 (%
$T_{c}\lesssim 29\,$K) to the one appropriate for Sm-1111 ($T_{c}\sim 50\,$K)%
\cite{AndersenBoeri2011}. Upon this tuning, our theoretical estimate for the
pairing temperature, the calculated scale of divergence for the fRG flow,
increases by a factor $\sim 3.$ In contrast with this, the energy scale for
SDW ordering for the undoped systems is only slightly increased. These differences between the two
systems are in qualitative agreement with the experimental phase diagrams.
It also agrees with previous works\cite{Kuroki2009} which showed that
regular tetrahedra have the highest superconducting energy scale. This
structural change is the main cause for the variation of the tuning
parameter. However, besides being able two make the system switch from a
high-pairing-scale five-pocket regime to a low-pairing-scale four-pocket
regime as previously pointed out\cite{Kuroki2009,ikeda,ThomaleMay2011}, the
tuning parameter also causes a distinct and relevant trend within the
five-pocket regime.
Hence our study reveals a major simplification regarding the modeling of the materials: it shows that  in the 8-band description the change of the pairing scale over 5- and 4-pocket regime can be described by adjusting a single parameter besides the band filling.

We have mentioned that simple measures like the density of states at the
Fermi level, or the values of one-loop diagrams, do not allow one to
understand the band-structure effects on the pairing scale in simple terms.
On the other hand, we have seen that the effect correlates positively with
the change of the Fermi-surface shape and velocities, but less with the
change of the orbital composition of the Fermi surface. The difficulty in
identifying a simple reason for the increase of the critical scales mainly
points to the complexity of the interplay of spin- and pairing fluctuations
in such multi-band systems. Nevertheless, we can associate the
scale-increase for the superconducting pairing with a single band structure
parameter, whose behavior can be understood from the structural change,
namely the pnictogen height.

In our view, these results and other previous works\cite%
{Kuroki2009,Wang2010,AndersenBoeri2011,ThomaleMay2011}, on material trends
on iron-arsenide models are the first steps towards a more ambitious program of
correlating theoretical results on different iron arsenide systems with
non-universal experimental findings. This serves two goals: first this
research helps us to understand how quantitative and how robust our present
treatments of the many-body physics in these systems are, and second the
so-obtained understanding may be very useful for the development of
materials with deliberately tailored, superior properties.

\section*{Acknowledgments}

We acknowledge financial support through the DFG priority program SPP1458 on
iron pnictides and the DFG research unit FOR723 on fRG methods. RT has
been supported by the European Research Council through ERC-StG-2013-336012.

\bibliography{lit}

\end{document}